\documentclass[preprint]{revtex4}

\usepackage{epsfig,latexsym,amssymb}
\usepackage{latexsym}
\usepackage{amssymb}
\usepackage{amsfonts}
\usepackage{comment}
\usepackage{graphicx}
\usepackage{verbatim}

\newcommand{\be}{\begin{equation}}
\newcommand{\ee}{\end{equation}}
\newcommand{\bea}{\begin{eqnarray}}
\newcommand{\eea}{\end{eqnarray}}

\begin{document}

\title{Probing Dark Matter Axions using the Hyperfine Structure Splitting of Hydrogen Atoms}

\author{Qiaoli Yang}
\email{qiaoliyang@jnu.edu.cn}
\author{Shiqin Dong}

\affiliation{Department of Physics and Siyuan Laboratory, Jinan University, Guangzhou 510632, China}

\begin{abstract}
QCD axions can be a substantial part of dark matter if their mass $m_a\sim10^{-5}$eV. Since the axions were created by the misalignment mechanism, their local energy spectrum density is large. Consequently, the axion-induced atomic transition rate is enhanced if the atomic energy gap matches the axion mass. The hyperfine splitting between the spin 0 singlet ground state and the spin 1 triplet state of hydrogen is $0.59\times10^{-5}$eV, which is close to the preferred mass of dark matter axions. With an energy gap adjustment by applying a weak Zeeman magnetic field, dark matter axions can induce atomic hydrogen transitions. Furthermore, because the total spins of the hydrogen triplet and singlet differ, the axion-induced transitions are detectable by a Stern--Gerlach apparatus or a sensitive magnetic field detector. A potential realization of the proposed scheme can be similar to existing hydrogen masers.
\end{abstract}

\date{\today}
\maketitle

\newpage

\section{Introduction}
The existence of cold dark matter in our universe is generally accepted due to abundant astrophysical and cosmological evidence. Observations indicate that the energy density in our universe is dominated by 73\% of dark energy followed by approximately 23\% of dark matter. The ordinary matter made of the standard model particles accounts for less than 4\%. Although the existence of dark matter is well established, its particular properties remain unknown.

The strong CP problem is one of the most important puzzles in particle physics today. Among various possible solutions, the Peccei--Quinn (PQ) theory \cite{Peccei:1977hh, Peccei:1977ur} suggests adding a global U(1) symmetry (the PQ symmetry) to the Lagrangian. The symmetry is broken both spontaneously and explicitly, which gives rise to a pseudo Nambu--Goldstone boson now known as the quantum chromodynamics (QCD) axion
\cite{Weinberg:1977ma, Wilczek:1977pj,vysotsky, Kim:1979if,Shifman:1979if,Zhitnitsky:1980tq,Dine:1981rt,Davidson:1981zd}. Interestingly, the symmetry breaking process naturally leads to a cosmological relic abundance of cold axions \cite{Preskill:1982cy,Abbott:1982af,Dine:1982ah,Sikivie:1982qv,Ipser:1983mw,berezhiani1985,Sikivie:2006ni,Svrcek:2006yi}. Therefore, axions are a natural candidate for cold dark matter. This symmetry-breaking creation process is often referred to as the misalignment mechanism.

By assuming that the PQ symmetry broke after inflation and the resulting axions constituted the majority of dark matter (DM), an important DM axion window is the PQ scale $f_a\sim 10^{11}$GeV and the axion mass $m\sim 10^{-5}$eV \cite{Sikivie:2006ni,Svrcek:2006yi,Hertzberg:2008wr,Sikivie:2009qn,Marsh:2015xka}. This is often called the classical window. There is an additional anthropic window for QCD DM axions if the PQ symmetry was broken before cosmological inflation. In the latter scenario, the anthropic selection resulted in a small initial misalignment angle; consequently, the PQ scale can be much larger than the prediction in the classical window. This scenario is constrained by cosmic microwave background (CMB) observations and is consistent with the low-scale inflation model \cite{Hertzberg:2008wr}. Typically, if $H_I<10^{10}$GeV, $f_a\gtrsim 10^{14}$GeV and $m_a\lesssim 10^{-7}$eV. Some recent studies, such as \cite{Graham:2018jyp,Co:2019jts,Sokolov:2022fvs}, relax these constraints. If one also considers axion-like particle dark matter, the theoretical mass range is even larger.

There are many proposed and ongoing experimental studies searching for axions due to the uncertainties of axion mass \cite{Sikivie:1983ip,Sikivie:1985yu,DePanfilis:1987dk,Hagmann:1990tj,Ehret:2010mh,Tam:2011kw,Wouters:2013iya,Graham:2013gfa,Budker:2013hfa,Sikivie:2013laa,Stadnik:2013raa,Ayala:2014pea,Rybka:2014cya,Sikivie:2014lha,Santamaria:2015gro,TheMADMAXWorkingGroup:2016hpc,Barbieri:2016vwg,Kahn:2016aff,Yang:2016zaz,Brubaker:2016ktl,Anastassopoulos:2017ftl,Abel:2017rtm,Graham:2017ivz,McAllister:2017lkb,Akerib:2017uem,Du:2018uak,Marsh:2018dlj,Zhong:2018rsr,Lawson:2019brd,Ouellet:2018beu,Arza:2019nta} etc. In this paper, we propose to use splitting of the $1S$ state of hydrogen atoms to probe the DM axions. Axions couple to fermions. Because of the high energy spectrum density of the DM axions, resonant atomic transitions are greatly enhanced compared to the nonresonant effects. This phenomenon has been considered in \cite{Sikivie:2014lha} in a general way, and \cite{Santamaria:2015gro} later suggested using molecular oxygen to search axions with a mass of approximately 10$^{-3}$eV. Interestingly, the hyperfine splitting of the hydrogen $1S$ state is $0.59\times 10^{-5}$eV (see FIG.1); therefore, it could be matched with the preferred dark matter QCD axion mass by applying a small external magnetic field. The magnetic moment of the hydrogen atoms can be detected as the sign of atomic transitions. Because a flip of either the proton or the electron causes the transition of the states, this scheme works for the KSVZ axions as well. Thus, using the hyperfine splitting of atomic hydrogen allows us to simultaneously probe both the axion-electron and axion-proton couplings with a single transition compared to just the axion-electron couplings in earlier proposals. The induced quantum transitions could be counted with a Stern-Gerlach apparatus or a maser-like device (see FIG. 3,4,5). The anthropic window of QCD axions or axion-like particles can also be explored by splitting the 1S triplet state (see FIG. 2).

\section{QCD axion mass considerations}
If the QCD axions were created by the misalignment mechanism in the early universe and compose the majority of dark matter, the preferred mass range can be theoretically constrained. Certainly, some mechanisms, such as \cite{Co:2019jts}, could result in a larger mass window.

After the Peccei--Quinn symmetry breaking, the equation of motion of the axion field $a$ in the Friedmann-Robertson-Walker (FRW) universe is
\be
\partial^2_t a+3H\partial_t a-{1\over R^2}\nabla ^2a+\partial_a V(a)=0~~,
\ee
where $R$ is the scale factor, $H=\dot R/R$ is the Hubble parameter, and $V(a)$ is the potential of the axion field. The potential depends on the temperature $T$ of the background and can be written as follows:
\be
V(a)\approx f_a^2m_a^2(T)[1-{\rm cos}({a\over f_a})]~~.
\ee
When $T\gtrsim \Lambda_Q$, $m_a(T)$ is
\be
m_a(T)\approx m_0b({\Lambda_Q\over T})^4~~,
\ee
where $\Lambda_Q\sim 200$MeV, and $b\sim {\cal O}(0.01)$ depending on the particular axion models. When $T\lesssim \Lambda_Q$, the axion mass is
\be
m_0\approx 6\times 10^{-5}{\rm eV}({10^{11}{\rm GeV}\over f_a})~~.
\ee

Because the Hubble parameter $H$ was large in the early universe, the field potential $V$ was negligible. Consequently, the initial misalignment angle $\theta_0=a_0/f_a$ was frozen. The axion field started to oscillate when the Hubble parameter dropped to $H\approx m_a(T_{osc})/3$, at which point the field potential was no longer negligible. Thus, the axion energy density at that time was
\be
\bar \rho_a\sim {1\over2}m_a(T_{osc})^2<\theta_0^2>f_a^2~~,
\ee
where $<\theta_0^2>$ is averaged over horizon so it gives rise to an order of one number. Subsequently, the energy density decreased as matter. Assuming the axions are the major part of dark matter, the implied parameter window is
\bea
f_a\sim 10^{11}{\rm GeV}\nonumber\\
m_a\sim 10^{-5}{\rm eV}.
\eea

If PQ symmetry breaking occurred before inflation, due to anthropic selection, $f_a$ could be significantly larger than in the previous scenario. The presence of the axion field during inflation generated isocurvature perturbations, which are constrained by the CMB \cite{Hertzberg:2008wr}, but recent studies, \cite{Graham:2018jyp}, etc., showed some relaxations. A possible DM QCD axion window for the low-scale inflation scenario, $H_I\lesssim 10^{10}$GeV, is
\bea
f_a\gtrsim 10^{14}{\rm GeV}\nonumber\\
m_a\lesssim 10^{-7}{\rm eV}.
\eea
Other production mechanisms and axion-like particles generally could have a much more relaxed mass constraint, but these two windows draw great interest due to their elegance and simplicity.

\section{Dark matter axion-induced quantum transitions}

\begin{figure}
\begin{center}
\includegraphics[width=0.5\textwidth]{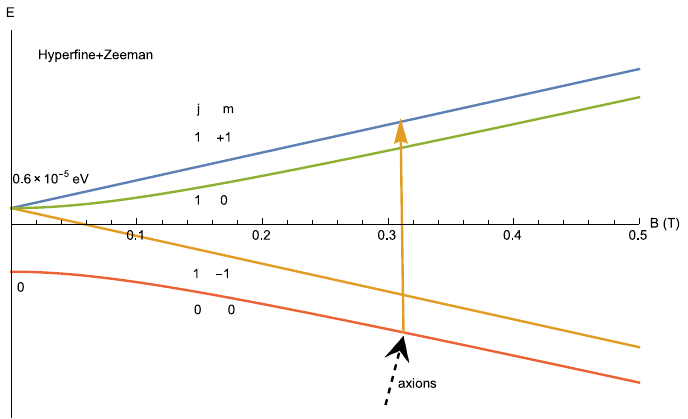}
\caption{Splitting of the hydrogen $1S$ state. In the classical window, the $|0,0>\to |1,1>$ transition is suitable for axion detection.}
\end{center}
\end{figure}

\begin{figure}
\begin{center}
\includegraphics[width=0.5\textwidth]{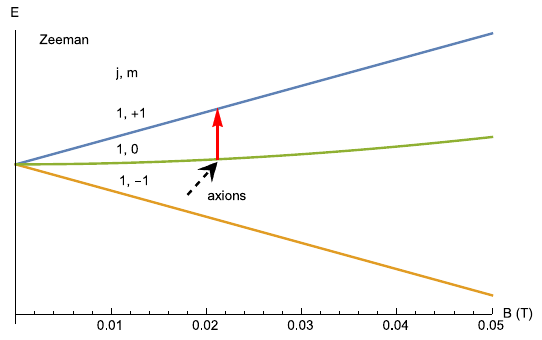}
\caption{Splitting of the hydrogen $1S$ triplet state. In the anthropic window, the $|1,0>\to|1,1>$ transition is suitable for axion detection.}
\end{center}
\end{figure}

At the laboratory scale, the dark matter axions can be considered free steaming. Therefore, they satisfy the Klein--Gordon equation:
\be
(\partial^2+m_a^2)a=0~~.
\ee
In addition, because the cold dark matter particles are nonrelativistic, the axion field can be written as \cite{Foster:2017hbq}:
\bea
a(x)&\approx& a_0\sum_j\alpha_j{\sqrt{f(v_j)\Delta v}}\nonumber \\
&\times&{\rm cos}(-m_at-{m_a\over 2}v_j^2t+m_a \vec v_j\cdot  \vec x+\phi_j)~~,
\eea
where $\alpha_j$ is a random number of the Rayleigh distribution $P(\alpha_j)=\alpha_j e^{-\alpha_j/2}$, $f(v_j)$ is the local DM speed distribution, $v_j\ll c$ is the local velocity relative to the laboratory, and $\phi_j$ is a phase factor. $f(v_j)$ is different depending on the particular halo models but is generally very sharp, so the DM axions can effectively be considered a mono-frequency field with a small frequency spread in experimental searches.

The averaged field strength: $\bar a_0\approx\sqrt{2\rho_{\rm CDM}}/m_a$,
where $\rho_{\rm CDM}\approx 1$GeV/cm$^3$ is the local dark-matter energy density. The local axion velocity $v\approx \bar v_j$ depends on the dark matter halo structure and the relative position of Earth. Assuming that the dark matter particles did not lose energy during the formation of galactic halos, their speed can be assumed to be approximately equal to the speed of the Sun, i.e., $v\sim 10^{-3}c$.

The local dark matter energy spectrum density is
\be
I_a={\rho_{CDM}\over (1/2)m_a\delta v^2}~~,
\ee
where $\delta v$ is the velocity spread. The typical estimation is $\delta v\sim 2\sqrt {T_a/M}\sim 10^{-3}$c, where $T_a$ is the effective dark matter temperature. Some authors suggest a lower dark matter temperature, which leads to $\delta v\sim 10^{-7}$c \cite{Armendariz-Picon:2013jej}.

The axions couple to electrons and protons via:
\be
{\cal L}_{int}=-\sum{g_f\over f_a}\partial_\mu a\bar \psi_f \gamma^\mu\gamma^5\psi_f \nonumber~~,
\ee
where $\psi$ is the fermion field. $f=e,p$ refers to electrons and protons, respectively. In atoms, electrons and protons can be regarded as nonrelativistic. Therefore,
\be
H_{int}={1\over f_a} \sum g_f(\partial_t a {\vec p_{f}\cdot \vec \sigma_{f}\over m_{f}}+\vec \sigma_{f}\cdot \vec \nabla a)
\label{int}
\ee
where $p_f$ and $\sigma_f$ are the momentum operator and the spin operator of the fermions, respectively. For the atomic transitions, the first term is subdominated compared to the second term. One could find this by exploring the commutator $\vec p_f=im_f[H, r_f]$, where H is the atomic Hamiltonian; Please see \cite{Stadnik:2013raa} Eq.(11)-(14) for details. Because $\langle A|\vec p\cdot \vec \sigma |B\rangle\propto (E_A-E_B)\langle A| \vec r\cdot \vec \sigma|B\rangle  m_f$, the first term in the bracket of Eq.(\ref{int}) is proportional to $m_a^2a_0\bar r$, where $\bar r\sim 10^{-11}$m is the Bohr radius. Thus, $m_a^2a_0\bar r<10^{-10}m_aa_0$ when $m_a<10^{-5}$eV. The second term is proportional to $v m_aa_0>10^{-3}m_aa_0$. Therefore, the first term can be dropped.

In addition, the wavelength of the axions is $\lambda=2\pi m_a^{-1}$, which is much larger than the Bohr radius $\bar r$ of atoms. Consequently, Eq.(\ref{int}) becomes
\be
H_{int}\approx\sum{g_f\over f_a} m_a a_0 \vec \sigma_f\cdot \vec v{\rm sin}(\omega_at)~~,
\ee
where $\omega_a=m_a(1+v^2/2)$ is the energy of the axions. When the energy gap between the atomic states matches the energy of the axions, the induced transition rate is
\be
R={ \pi\over f_a^2}|\sum g_f<f|(\vec v\cdot \vec \sigma_f)|i>|^2{I_a}~~.
\label{rate}
\ee
Eq.(\ref{rate}) can only be applied when the initial atomic state $|i>$ has a lifetime longer than the axion oscillation time $2\pi/m_a$, which is true for the hydrogen 1S state. The resonant transitions also require a match between the transferred energy and the atomic energy gap, which can be realized by using the Zeeman effect (see Figs. 1, 2). For the classical window, let us consider the energy splitting between the singlet state and the triplet state:
\bea
&\Delta E&\approx 2\mu_B B+5.9\times10^{-6}{\rm eV} \\ \nonumber
&=&({\rm B/Tesla})\times11.6\times10^{-5}{\rm eV}+5.9\times10^{-6}{\rm eV}~~,
\eea
For example, the external field should be approximately 0.05 T for $m_a\sim 10^{-5}$eV. For the anthropic window, let us consider the Zeeman splitting of the triplet:
\bea
\Delta E'&\approx& 3\times10^{-6}{\rm eV}+2.6\times10^{-5}{\rm eV\times (B/Tesla)} \nonumber\\
&-&3\times10^{-6}{\rm eV}\sqrt{1+76.5({\rm B/ Tesla})^2}~~.
\eea
A recent work estimated that $f_a\sim 10^{15}$GeV \cite{Gao:2019tqt}. Then, the external field should be approximately 0.001 T. Note that when the Zeeman field is very weak, the $|1,1>$ and $|1,-1>$ splitting is almost equal in energy; thus, $|1,0>$ transitions to both states due to axion DM energy spectrum spreading. The transition rate Eq.(\ref{rate}) is then doubled for the anthropic window searches.

The mass range that can be scanned is limited by the available strength of the Zeeman field. For current technology, the classical window can be fully covered, and the anthropic window is limited by nT, which is approximately $10^{-14}$eV$\sim10^{-7}$eV.

The axion models predict that at least one of $g_e$ and $g_p$ is on the order of one. Therefore, $|\sum g_f<f|(\vec v \cdot \vec \sigma_f)|i>|^2\sim v^2$ for $|j_i,0>\to |j_f,1>$ transitions. The event rate is then
\bea
&~&~NR=N{\pi \over f_a^2m_a}({v\over \delta v})^2\rho_{\rm CDM}\\ \nonumber
&~&=N\times3.7\times10^{-25}({ v\over \delta v})^2{({t\over{\rm s}}) ({m_a\over {\rm 10^{-5}eV}})^{-1}({f_a\over 10^{11}{\rm GeV}})^{-2}}~.
\eea
For $\delta v\sim 10^{-3}$c, $m_a\sim 10^{-5}$eV, $f_a\sim 10^{11}$GeV, and $N\sim 1$mole, the event rate is 2.0 s$^{-1}$. If $\delta v\sim 10^{-7}$c \cite{Armendariz-Picon:2013jej}, only $10^{-8}$ mole atoms are required to achieve a similar event rate. We see that a smaller mass can partially nullify the increase in $f_a$.

The major source of noise using this approach is the thermal excitation of atomic states; then, the optimal temperature $T_o$ satisfies
\be
N_{thermal}={1\over{\rm e}^{\Delta E/k_B T_{o}}-1}<1~~,
\ee
where $\Delta E\approx m_a$. If $m_a\approx 10^{-6}$eV, we have $T_{o}\approx 20$mK. When the temperature is higher than the optimal temperature, the detection requires a longer integration time that satisfies:
\be
{R/ R_n^{1/2}}\times(N\times t)^{1/2}>3,
\ee
where $R_n$ is the thermal-induced transition rate.

\section{Possible Experimental Set-ups}
The setup of the experiment can be very similar to the original experiment demonstrating the Zeeman splitting transitions (see Fig.3), where in addition to the Zeeman field, a small varying magnetic field that varies at GHz was applied. A major difference is that in the proposed experiment, the dark matter axions stimulate the transitions; therefore, a varying field is not needed. On the other hand, a hydrogen maser-like device with some modifications could be ideal for the experiment (see Fig.4). The molecular hydrogen is dissociated in the discharge bulb into individual hydrogen atoms, and the beam passes the state selector, so the $m=0$ states are selected. The selected atoms can be stored and cooled in a storage bulb, which is widely used in maser experiments \cite{Kleppner:1965zz}. The inside of the storage bulb is typically coated with Teflon, so many collisions of the atoms with the wall do not change the atomic state. Current bulb technology can achieve an order of years lifetime of an atomic state. The atomic hydrogens could be further cooled by contacting the low-temperature bulb wall. The atomic energy gaps are adjusted by an external magnetic field. When the energy gap matches the axion mass, the atoms will be stimulated to the $m=1$ state, which can be detected. The $m=0$ and $m=1$ states are relatively easy to distinguish; thus, we expect that the detection efficiency could be very high. In addition, axion-induced atomic transitions occur when the atomic state energy gap roughly overlaps with the peak part of the axion energy spectrum density $I_a(\omega_a)$. The effective integration time at each axion frequency pin is approximately several seconds, so during that time period, $\delta t\sim$seconds, the change in the Earth's velocity is relatively small, of order $O(\delta t/1 year)$. Thus, the earth-induced axion energy shift is small compared to the detector bandwidth at a given time, which should simplify the scanning scheme.

\emph{Some Maser-like device considerations}:

\emph{Atom Escapes}: The atoms inside a bulb eventually escape through the entering hole. The escape rate $\gamma_0$ of atoms from a bulb is $\gamma_0=\bar v_H A_e/(4KV_b)$ \cite{Kleppner:1962zz}, where $\bar v_H$ is the atomic mean velocity, $A_e$ is the total escape area, $K$ is a form factor order of one depending on the geometry of the bulb, and $V_b$ is the volume of the bulb. The escape rate is very low assuming the optimal temperature, with a typical spherical Teflon bulb of ${\cal O}$(meters) in diameter and a thin hole ${\cal O}$(mm) in diameter. The typical escape is approximately 1 sec$^{-1}$, so with a small amount of fresh hydrogen atom supply, a large amount of atoms can be stored inside the bulb.

\emph{Atom Collisions with the Wall}: The atoms inside the bulb make random collisions with the Teflon-coated wall. The existing experiments show that the atomic states are not seriously changed due to collisions, and these atoms spend most of the time in free motion in the bulb \cite{Ramsey:1990zz}. The collisions can be categorized into two categories: adiabatic and nonadiabatic. The adiabatic collisions do not change the atomic states but change the atomic energy level slightly, which has no important consequences in the axion searches, as the energy shift occurs very shortly. The nonadiabatic collisions or chemical reactions with the surface change the atomic states, which could contribute to the experimental noise. Fortunately, this type of collision only occurs when the incident atom possesses a kinetic energy equal to or exceeding the activation energy of the reaction. The rate $\gamma_1$ is \cite{Kleppner:1962zz}
\bea
\gamma_1={2\bar v_HP\over \pi^{1/2}l}e^{(-E_a/kT)}~,
\eea
where $\bar v_H=\sqrt{3kT/m}$ is the rms velocity of atoms, $l$ is the mean distance between collisions, $P$ is a factor counting the reaction percentage of collisions, which is usually taken as ${\cal O}(0.1)$, and $T$ is the storage bulb temperature. The activation energy $E_a$ depends on the chemical reactions of the atom and the bulb wall \cite{Kleppner:1962zz}, which is typically much higher than the thermal energy of atoms at the proposed experimental temperature.

\emph{Atomic State Relaxations}: Ideally, if the magnetic field in the storage bulb can be perfectly uniform, it will not induce atomic transitions. However, if there is a nonuniform part of the magnetic field, atomic motion induces transitions. Therefore, some counts should be taken if the atom leakage is large. The $|0,0>$ state is much less perturbed by this phenomenon, while the $|1,0>$ state contributions dominate. The relaxation rate depends on the nonuniformity of the field, the storage bulb geometry, and the atomic speed $\bar v_H$, which is small after reaching the proposed atomic temperature. In the limit of a low strength Zeeman magnetic field, which is the case in the proposed setup, this relaxation rate is typically much smaller than the axion-induced transition rate $\sim\cal O$(Hz).

In addition, if there are different spin states in the storage bulb, hydrogen-hydrogen spin-exchange collisions can contribute to atomic state relaxation. The experimental and theoretical analyses predict that this effect is density related \cite{Kleppner:1962zz}. This decay rate is $\propto 10^{-10}N$sec$^{-1}$, so several stages of the state selection filter are preferable to filter the different spin states before they enter the bulb.

\emph{Doppler Effects}: The Doppler shifts due to atomic thermal velocities can broaden the resonance width. The typical atomic thermal velocity is approximately $\bar v_H\sim 10^{-8}$c under the optimal temperature $T_o$; thus, the resulting broadening should be much smaller than the spectrum width of the DM axion itself. Note that the hydrogen atoms are confined in a region of almost constant phase space because the adiabatic atomic collisions with the wall do not change the atomic speed. Nevertheless, this Doppler effect may slightly reduce the transition rate at the center of the resonance line. One may define a quality factor of the hydrogen atoms in the bulb: $Q_H=\omega/\delta \omega$, where $\delta \omega$ is due to the Doppler broadening and $\omega$ is the center of the transition frequency. The value of $Q_H$ depends on the particular experimental setup but remains constant during the integration time. As long as it is higher than the axion DM quality factor $Q_a=1/\delta v_a^2\sim 10^6$, the sensitivity will not be reduced.

\section{Scanning and sensitivity}
\begin{figure}
\begin{center}
\includegraphics[width=0.45\textwidth]{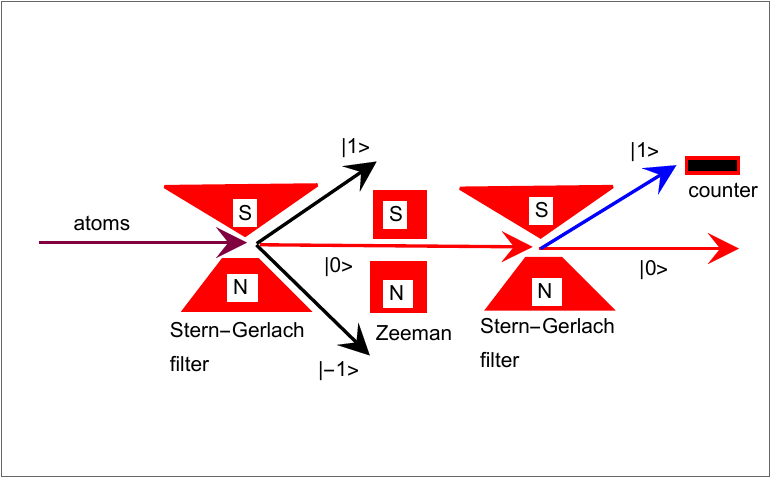}
\caption{Possible setup of the proposed detection scheme. The cold atoms enter the first Stern--Gerlach apparatus in which the $m_j\neq 0$ states are filtered. Subsequently, the $m_j=0$ atoms go through a Zeeman effect region where their atomic energy gaps are adjusted to match the axion mass. A small portion of the atoms is resonantly excited to the $m_j=1$ state, which is deflected by the second Stern-Gerlach apparatus toward the counter. The apparatus sensitivity depends on the number of atoms $N$ present in the Zeeman effect region and the scanning bandwidth $\Delta f$.}
\end{center}
\end{figure}

\begin{figure}
\begin{center}
\includegraphics[width=0.45\textwidth]{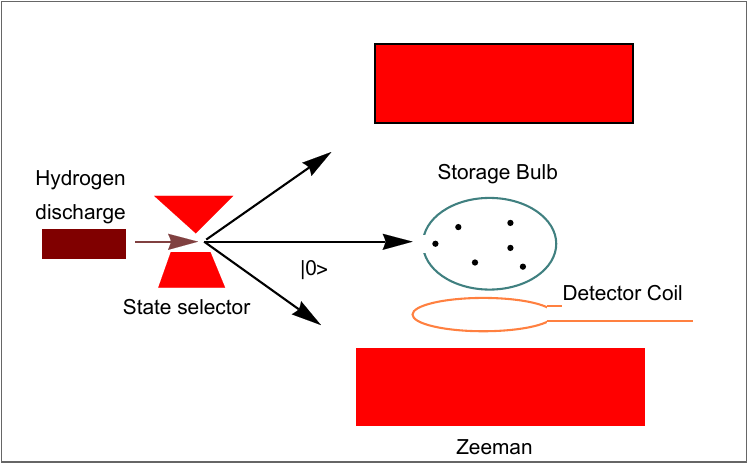}
\caption{Another possible setup of the proposed detection scheme, which is similar to the hydrogen masers. The cold atoms go through a magnetic gate that allows only $m_j= 0$ states to pass. Subsequently, the atoms enter a storage bulb surrounded by a tuneable Zeeman effect magnet. Once inside the bulb, the atomic energy gaps are matched with the axion mass. Some atoms are resonantly excited to the $m_j=1$ state, which is detectable by the magnetic field detector.}
\end{center}
\end{figure}

\begin{figure}
\begin{center}
\includegraphics[width=0.45\textwidth]{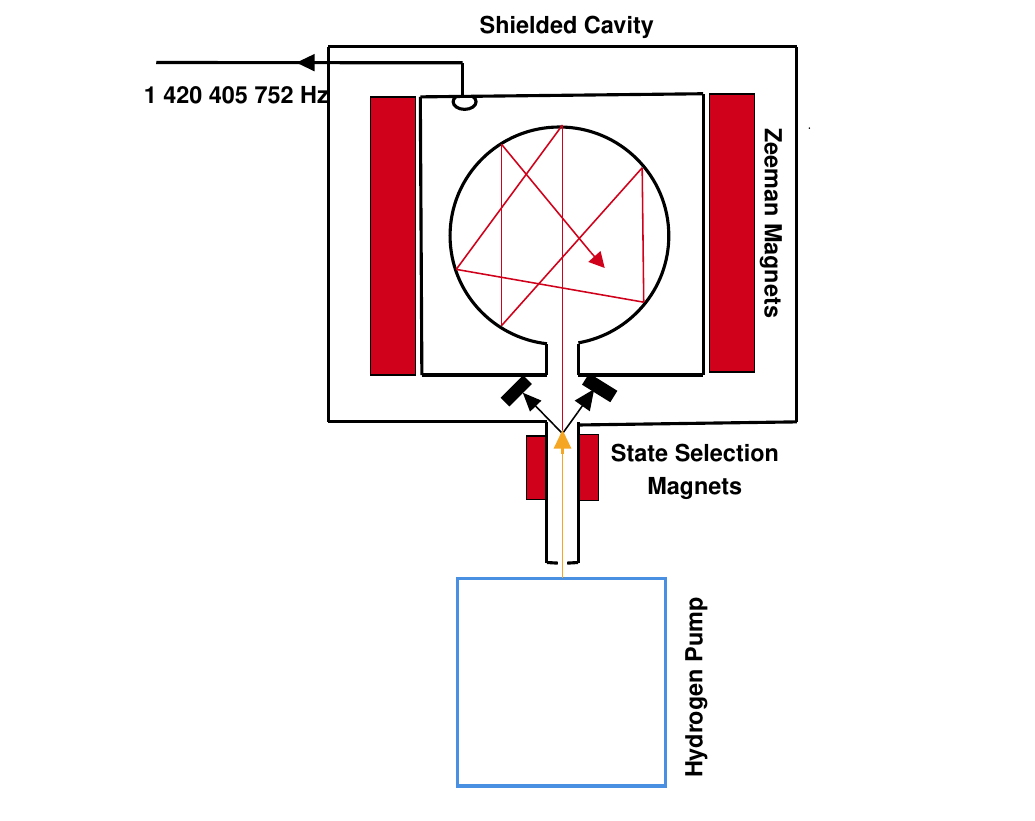}
\caption{An illustration of a typical hydrogen maser. With some modifications, such as adding Zeeman magnets, the scheme of Fig. 4 can be realized.}
\end{center}
\end{figure}

Because the exact value of the axion mass is unknown, one needs to scan the interested mass range. This can be done by turning the Zeeman magnetic field $B$. Assuming to scan $\Delta f\equiv \Delta m/2\pi$ in a single working year, the magnetic field is tuned as
\bea
R_{\rm scaning}={\Delta f\over t_{exp}}=({\Delta f\over {\rm GHz}})\times31.71{\rm Hz/s}~~.
\eea
To cover approximately ${\cal O}(10^{-5})$eV mass range, one needs to scan approximately several GHz. To cover a portion of the anthropic axion mass window, e.g., $m_a\sim 10^{-8}$eV, the scanning bandwidth is approximately several MHz. Since the bandwidth of the DM axion wave is $\Delta f_a={1\over 2}m_a\delta v^2$, the effective integration time $t_{int}=\Delta f_a/R_{scaning}$ is
\be
t_{int}=3.8\times10^{7}{\rm s}({m_a\over 10^{-5}{\rm eV}})\delta v^2({{\rm GHz}\over\Delta f})~~.
\ee
Assuming operating at the optimal temperature and high confidence event detection, $3\sigma$ detection requires $NRt_{int}\geq 3$, which leads to:
\bea
f_a\leq 2.82\times10^{11}{\rm GeV}\sqrt{{{\rm GHz}\over\Delta f}}\sqrt{{N\over {\rm mole}}}~~.
\eea
An inefficiency of the event detection, e.g., only x\% events detected, will slightly reduce the sensitivity as $\sqrt{x\%}$~. To search the anthropic axion DM, say $f_a\sim 10^{15}$GeV and $m_a\sim 10^{-8}$eV, with 10\% frequency range $\Delta f \approx0.1*10^{-8}{\rm eV}/2\pi=0.24$MHz covered in one year, $N\sim 10^3$ mole. To search the classical axion DM, $f_a\sim 10^{12}{\rm GeV}$ and $m_a\sim 10^{-5}$eV, with 10\% frequency range $\Delta f\approx 0.1*10^{-5}{\rm eV}/2\pi=0.24$GHz covered in one year, $N\sim 1$ mole. Please see FIG. 6 for the sensitivity estimation.

\begin{figure}
\begin{center}
\includegraphics[width=0.5\textwidth]{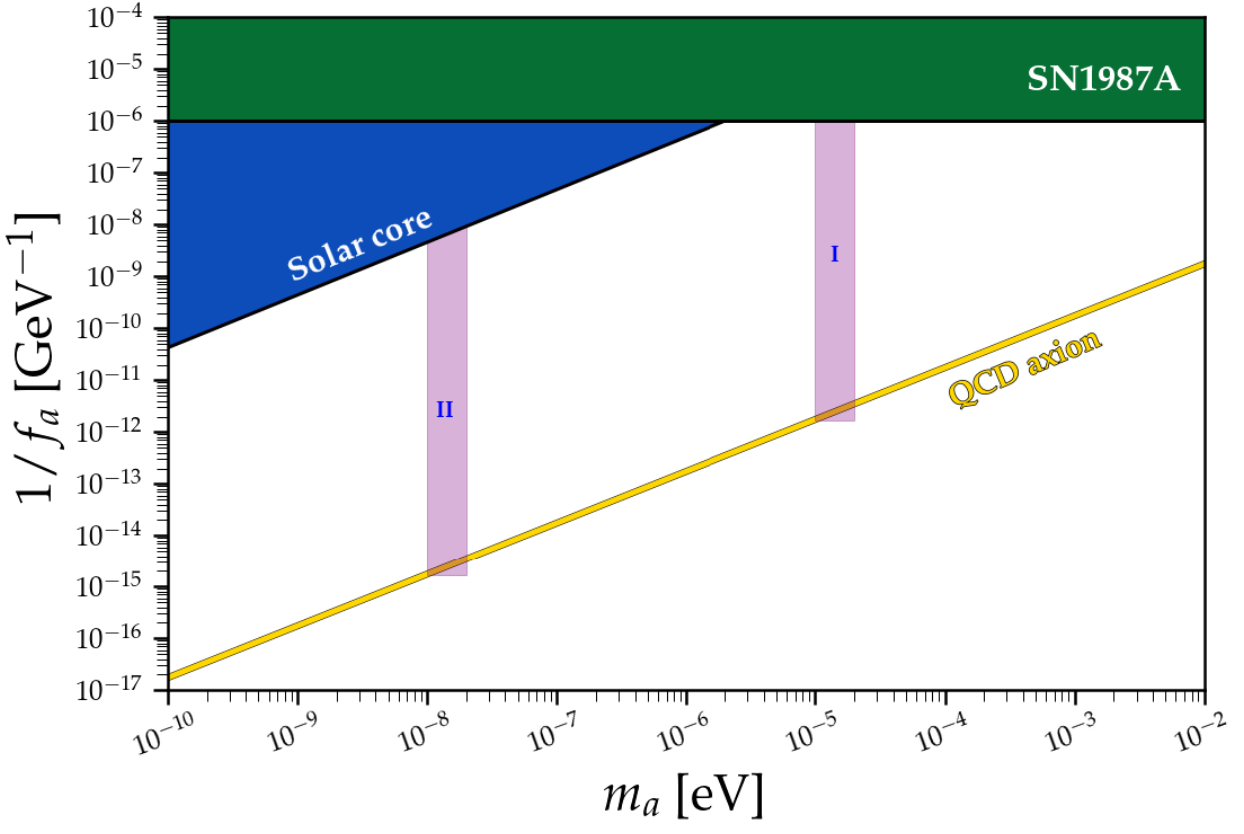}
\caption{Constraints of the proposed scheme with one year data-taking time. The mass range that can be covered by the proposed method is determined by the available Zeeman field strength. To cover region I, $m_a\in [10^{-5}{\rm eV},2\times 10^{-5}{\rm eV}]$, 0.15 Tesla is needed. Assuming one covers 10\% of the region with $1$mole of H, the sensitivity is $f_a=5.8*10^{11}$GeV. For region II, $m_a\in[10^{-8}{\rm eV},2\times 10^{-8}{\rm eV}]$, 0.002 T is needed. Assuming one covers 10\% of the region with $10^{3}$mole of H, the sensitivity is $f_a=5.8*10^{14}$GeV. The yellow band defines the QCD axion parameter space. The solid-color filled regions, SN1987A and Solar Core, have been excluded by experiments and observations.}
\end{center}
\end{figure}

\section{Conclusion}
The QCD axion is a well-motivated dark matter candidate. It could provide information about the ultrahigh-energy new physics, $f_a\gtrsim 10^{11}$GeV. It may also generate rich cosmological phenomena, for example, implications for the inflation Hubble scale $H_I$. Thus, it is crucial and promising to pursue direct laboratory searches of axions or axion-like particles. Nevertheless, the construction of experimental probes is rather challenging because the axions only weakly couple to the standard model particles, and the energy carried by each axion is very small. Therefore, the axion interactions are weak in terms of energy.

Fortunately, DM axions are highly occupied in the phase space. Hence, their interactions with fermions can be enhanced by the resonant effect. Hydrogen atoms could be a perfect target owing to their atomic energy gaps being close to the possible mass windows of the DM axions. Since the amount of energy transferred per event is very small, measuring the spin instead of the transferred energy might be more straightforward.

\section*{Acknowledgments}
We are grateful for useful discussions with Pierre Sikvie, Giovanni Carugno, Nick Huston and Man Jiao. This work is supported by the NSFC under grant no.11875148, no.12150010 and is, in part, supported by the Gordon and Betty Moore Foundation under grant GBMF6210.

\end{document}